\global\def\draftcontrol{0}

   \def\versionno{ csc can't be saved }

\catcode`\@=11

\expandafter\ifx\csname draftcontrol\endcsname\relax\global\def\draftcontrol{0}
\fi

{\count255=\time\divide\count255 by 60
\xdef\hourmin{\number\count255}
\multiply\count255 by-60\advance\count255 by\time
\xdef\hourmin{\hourmin:\ifnum\count255<10 0\fi\the\count255}}
\def\draftdate{\number\month/\number\day/\number\year\ \ \ \hourmin }

\newcommand\makepapertitle{\par
  \begingroup
    \renewcommand\thefootnote{\@fnsymbol\c@footnote}%
    \def\@makefnmark{\rlap{\@textsuperscript{\normalfont\@thefnmark}}}%
    \long\def\@makefntext##1{\parindent 1em\noindent
            \hb@xt@1.8em{%
                \hss\@textsuperscript{\normalfont\@thefnmark}}##1}%
     \newpage
     \global\@topnum\z@   
     \@makepapertitle
     \thispagestyle{empty}\@thanks
  \endgroup
  \setcounter{footnote}{0}%
  \global\let\thanks\relax
  \global\let\makepapertitle\relax
  \global\let\@makepapertitle\relax
  \global\let\@thanks\@empty
  \global\let\@author\@empty
  \global\let\@date\@empty
  \global\let\@title\@empty
  \global\let\title\relax
  \global\let\author\relax
  \global\let\date\relax
  \global\let\and\relax
  \def\version{\let\version\@version\@gobble}
}
\def\@makepapertitle{%
  \newpage
   \ifnum\draftcontrol=1 {}
   \version\versionno
   \vskip 3em%
   \else
   \hfill\hbox to 3cm {\parbox{4cm}{\@pubnum}\hss}%
   \vskip 3em%
   \fi
   \begin{center}%
   \let \footnote \thanks
     {\LARGE {\@title}}%
     \vskip 1.5em%
     {\normalsize
       \lineskip .5em%
       \begin{tabular}[t]{c}%
         \@author
       \end{tabular}\par}%
     \vskip 1.5em%
     {\@bstract}%
     \end{center}%
     \vskip 1.5em
     \@date%
   \par
}

\gdef\@pubnum{}
\def\pubnum#1{%
  \gdef\@pubnum{#1}}

\gdef\@bstract{}
\def\Abstract#1{%
  \gdef\@bstract{%
   \parbox{\textwidth-0pc}{%
   \centerline{\bf Abstract}\penalty1000%
\kern.2cm%
\noindent
\renewcommand\baselinestretch{1.0}%
{#1}}}
}

\def\ps@paper{\let\@mkboth\@gobbletwo%
     \ifnum\draftcontrol=1
    \def\@oddfoot{\hbox to \textwidth{\tiny \versionno \hfil\tiny\draftdate}%
    \hskip -\textwidth \hbox to \textwidth{\hfil\rm\thepage\hfil}}%
     \else\def\@oddfoot{\hbox to \textwidth{\hfil\rm\thepage\hfil}}
     \fi
     \let\@evenfoot\@oddfoot
}

\def\body{\clearpage
          \pagestyle{paper}
    }

\def\@version#1{\ifnum\draftcontrol=1
\typeout{}\typeout{#1}\typeout{}
\vskip3mm\centerline{\hbox{\fbox{\normalsize{\tt DRAFT -- #1 -- }
                   {\draftdate}}}}\vskip3mm
\fi}
\let\version\@version
\long\def\eqlabel#1{\ifnum\draftcontrol=1
                    \tag@false  
                    \tag*{(\theequation) \hbox to -0.2cm{\hspace{0cm}\small{#1}\hss}}
                    \refstepcounter{equation}
                    \edef\@currentlabel{\theequation}
                    \ltx@label{#1}          
                    \else
                    \label{#1}
                    \fi
                    }
\let\st@bibitem\@bibitem
\let\st@lbibitem\@lbibitem
\ifnum\draftcontrol=1
  \def\@bibitem#1{%
    \st@bibitem{#1}\a@@label{#1}\ignorespaces}
  \def\@lbibitem[#1]#2{%
    \st@lbibitem[#1]{#2}\a@@label{#2}\ignorespaces}
  \def\a@@label#1{%
    \gdef\a@lab{\smash{\normalfont\small#1}}
    \ifvmode
      \if@inlabel
        \global\setbox\@labels\hbox{%
          \llap{\a@lab\let\a@lab\relax
                \kern\@totalleftmargin\kern\marginparsep}%
          \box\@labels}%
      \fi
    \fi}
\fi

\documentclass[12pt,letterpaper]{article}

\usepackage{amsmath,amssymb,array,calc,epsfig,rotating,bm}
\usepackage[sort]{cite}
\usepackage{graphicx}
\usepackage{psfrag,verbatim}

\ifnum\draftcontrol=1
\fi

\renewcommand\baselinestretch{1.25}
\renewcommand\section{\@startsection {section}{1}{\z@}%
                                   {-3.5ex \@plus -1ex \@minus -.2ex}%
                                   {2.3ex \@plus.2ex}%
                                   {\normalfont\large\bfseries}}
\renewcommand\subsection{\@startsection{subsection}{2}{\z@}%
                                   {-3.25ex\@plus -1ex \@minus -.2ex}%
                                   {1.5ex \@plus .2ex}%
                                   {\normalfont\normalsize\bfseries}}
\renewcommand\subsubsection{\@startsection{subsubsection}{3}{\z@}%
                                   {-3.25ex\@plus -1ex \@minus -.2ex}%
                                   {1.5ex \@plus .2ex}%
                                   {\normalfont\normalsize\it}}
\renewcommand\paragraph{\@startsection{paragraph}{4}{\z@}%
                                   {-3.25ex\@plus -1ex \@minus -.2ex}%
                                   {1.5ex \@plus .2ex}%
                                   {\normalfont\normalsize\bf}}

\numberwithin{equation}{section}

\def\revise#1       {\raisebox{-0em}{\rule{3pt}{1em}}%
                     \marginpar{\raisebox{.5em}{\vrule width3pt\
                     \vrule width0pt height 0pt depth0.5em
                     \hbox to 0cm{\hspace{0cm}{%
                     \parbox[t]{4em}{\raggedright\footnotesize{#1}}}\hss}}}}

\newcommand\nxt[1]  {\\\fnxt#1}

\def\cale         {{\cal E}}
\def\calf         {{\cal F}}
\def\calg         {{\cal G}}

\def\caln         {{\cal N}}
\def\calo         {{\cal O}}

\def\del          {\partial}

\def\sqr#1#2{{\vcenter{\vbox{\hrule height.#2pt
 \hbox{\vrule width.#2pt height#1pt \kern#1pt
 \vrule width.#2pt}\hrule height.#2pt}}}}

\def\a{\alpha}

\newcommand{\hhh}{\mathbb{H}}

\newcommand{\beq}{\begin{equation}}
\newcommand{\eeq}{\end{equation}}
\newcommand{\beqa}{\begin{eqnarray}}
\newcommand{\eeqa}{\end{eqnarray}}
\newcommand{\beqar}{\begin{eqnarray*}}
\newcommand{\eeqar}{\end{eqnarray*}}

\renewcommand{\eqref}[1]{(\ref{#1})}

\newcommand{\ie}{{\it i.e.,}\ }

\def\a{\alpha}

\def\c{\chi}
\def\k{\kappa}

\catcode`\@=12

\begin{document}


\title{\bf Can the correlated stability conjecture be saved?}
\pubnum
{UWO-TH-11/3
}

\date{February 2011}

\author{
Alex Buchel$ ^{1,2}$ and Alexander Patrushev$ ^{1}$\\[0.4cm]
\it $ ^1$Department of Applied Mathematics\\
\it University of Western Ontario\\
\it London, Ontario N6A 5B7, Canada\\
\it $ ^2$Perimeter Institute for Theoretical Physics\\
\it Waterloo, Ontario N2J 2W9, Canada\\
}

\Abstract{
Correlated stability conjecture (CSC) proposed by Gubser and Mitra
\cite{gm1,gm2} linked the thermodynamic and classical (in)stabilities 
of black branes. In \cite{b} it was shown that the thermodynamic
instabilities, specifically the negative specific heat, indeed result
in the instabilities in the hydrodynamic spectrum of holographically
dual plasma excitations.  Counter-examples of CSC were presented in
the context of black branes with scalar hair undergoing a second-order
phase transition \cite{ccsc,bpcsc}. The latter translationary
invariant horizons have scalar hair, raising the question whether the
asymptotic parameters of the scalar hair can be appropriately
interpreted as additional charges leading to a generalization of the
thermodynamic stability criterion. In this paper we show that the
generalization of the thermodynamic stability criterion of this type
can not save CSC. We further present a simple statistical model which
makes it clear that thermodynamic and dynamical (in)stabilities
generically are not correlated.
}

\makepapertitle

\body

\version\versionno
\tableofcontents

\section{Generalized CSC}

A standard claim in classical thermodynamics\footnote{Assuming that the temperature 
is positive.} is that a system is thermodynamically stable if the Hessian $\hhh^\cale_{s,Q_A}$ of the energy 
density $\cale=\cale(s,Q_A)$ with respect to the entropy density $s$ and charges 
$Q_A\equiv\{Q_1,\cdots Q_n\}$, \ie
\begin{equation}
\hhh^\cale_{s,Q_A}\equiv \left(
\begin{array}{cc}
\frac{\del^2 \cale}{\del s^2} & \frac{\del^2\cale}{\del s\del Q_B}  \\
\frac{\del^2\cale}{\del Q_A\del s} & \frac{\del^2 \cale}{\del Q_A\del Q_B}  \end{array}
\right)\,,
\eqlabel{defhess}
\end{equation}
does not have negative eigenvalues. In the simplest case $n=0$, \ie no conserved charges, the thermodynamic stability 
implies that 
\begin{equation}
0\ <\ \frac{\del^2 \cale}{\del s^2}=\frac{T}{c_v}\,, 
\eqlabel{n0case}
\end{equation}
that is the specific heat $c_v$ is positive. In the context of gauge theory/string theory 
correspondence \cite{m9711} black holes with translationary invariant horizons in asymptotically anti-de-Sitter space-time 
are dual (equivalent) to equilibrium 
thermal states of certain strongly coupled systems. 
Thus, the above thermodynamic stability criterion should be 
directly applicable to black branes as well. The correlated stability conjecture (CSC) asserts that it is only 
when the Hessian \eqref{defhess}
for a given black brane geometry is positive, the spectrum of on-shell excitations in this background geometry 
is free from tachyons \cite{gm1,gm2}.  

In the simplest case, \ie the absence of the chemical potentials, one can 
trivially identify the classical instabilities of the thermodynamically 
unstable system \cite{b}.
Indeed, since the speed of sound waves squared in this case is
$c_s^2=\frac {s}{c_v}$, the thermodynamic instability of the system ($c_v<0$)
immediately implies that the hydrodynamic (sound) modes are classically 
unstable. There is no simple argument implying that thermodynamic stability of the system is 
enough to secure its classical stability; moreover, the instability link with the sound waves  
does not work in strongly coupled  R-charged $\caln=4$ supersymmetric Yang-Mills
plasma \cite{bpcsc} --- here, the speed of sound is always fixed to a 
conformal value $c_s^2=\frac 13$ even though there is an equilibrium 
branch with $c_v<0$.

In \cite{ccsc,bpcsc} it has demonstrated that, at least for a canonical interpretation of the black brane 
thermodynamics, the CSC is violated in the case of black branes with scalar hair that undergo a continuous 
phase transition. The dual gauge theory picture makes such violation almost self-evident. 
Indeed, in the vicinity of a continuous phase transition the condensate does not noticeably 
modify the thermodynamics, and thus should not affect the thermodynamic stability
of the system. On the other hand, the phase of the system with the higher 
free energy is expected to be classically unstable. The condensation of the tachyon 
should bring the system to the equilibrium phase with the lowest free energy.

The important qualifier for the above counter-examples is the {\it canonical interpretation} 
of the corresponding black brane thermodynamics. Specifically, the black branes considered 
have scalar hair and in the proper boundary (field theoretic) thermodynamic interpretation 
one has to keep  non-normalizable coefficients of the scalars fixed. The reason for this is that 
these non-normalizable coefficients
are dual to mass-scales in the boundary field theory. In thermodynamic stability analysis 
one naturally would like to keep microscopic mass scales in the field theory fixed. If one   
abandons the gauge/gravity analogy and considers black branes as thermal systems in 
higher dimensional general relativity, the motivation for keeping the 
asymptotic scalar hair parameters fixed is removed. It is an interesting question
as to whether these parameters might be treated as generalized charges in the context 
of thermodynamic stability of translationary invariant horizons in such a way that the CSC 
is validated\footnote{We would like to thank Barak Kol for raising this possibility.}.      
We argue here that CSC generalizations of these type are false. 

In the next section we present a simple statistical model in which the generalized thermodynamic  
and the dynamical (in)stabilities are not correlated. In section 3 we show that 
the exotic hairy black branes discussed in \cite{bpcsc,bp2,bp3}  
while classically unstable, are thermodynamically stable in the generalized manner outlined above.
In both cases the classical instabilities we identify are long-wavelength, provided, in the statistical model in section 
2, $\Lambda\ll T$, and for exotic hairy black branes one stays close to the phase transition.

\section{Counter-example to generalized CSC in statistical physics}

Consider a Landau-Ginsburg model with the following free energy density functional
\begin{equation}
\calf=-T^4+\Lambda^4 +\frac 12\left(\vec{\nabla}\phi(\vec{x})\right)^2-\frac 12 \Lambda^2\phi(\vec{x})^2
+\frac 14 \phi(\vec{x})^4\,, 
\eqlabel{lg}
\end{equation}
where $\Lambda$ is a mass-scale, and $\phi(\vec{x})$ is a dynamical scalar field.
For any temperature $T$, there are three equilibrium states of the system: 
one unstable $\ _u$ and two degenerate stable ones $\ _s$, 
\begin{equation}
\begin{split}
&\langle\phi(\vec{x})\rangle\bigg|_{unstable}=0\,,\qquad \Rightarrow\qquad \calf_u=-T^4+\Lambda^4\,,\\
&\langle\phi(\vec{x})\rangle^\pm\bigg|_{stable}=\pm\Lambda\,,\qquad \Rightarrow\qquad \calf_s^\pm=-T^4+\frac 34\Lambda^4\,.\\
\end{split}
\eqlabel{estates}
\end{equation}
In what follows we focus on the unstable equilibrium. Here, the energy density $\cale_u$
is given by 
\begin{equation}
\cale_u=\frac{3}{2^{8/3}}\ s^{4/3}+\Lambda^4\,,
\eqlabel{energy}
\end{equation}
where $s$ is the entropy density. 
It is straightforward to see that whether or not we treat the scale $\Lambda$ as a generalized charge $Q_A$
in the context of the thermodynamic stability (see \eqref{defhess}),
this classically unstable equilibrium is thermodynamically stable.  In other words, both 
Hessians $\hhh^{\cale_u}_s$ and $\hhh^{\cale_u}_{s,\Lambda}$ are positive.

Notice that since the energy $\cale_s$ of the stable equilibrium is 
\begin{equation}
\cale_s=\frac{3}{2^{8/3}}\ s^{4/3}+\frac 34\Lambda^4\,,
\eqlabel{energys}
\end{equation}
any other definition of the generalized charge $Q_A=f(\Lambda)$ would imply that the two equilibria 
$\ _u$ and $\ _s$ are simultaneously either thermodynamically stable or not\footnote{Clearly, we need to 
restrict definition of $Q_A$ so that $\ _s$ equilibrium is thermodynamically stable.}.
It might be possible to define a generalized charge $Q_A$ which depends both on $s$ and $\Lambda$, \ie $Q_A=f(s,\Lambda)$,
 so that $\ _s$ state is thermodynamically stable while the  $\ _u$ state is  thermodynamically unstable ---
it is not clear to us how to make such a definition universally for all statistical systems.

The model \eqref{lg} is probably the simplest example which clearly demonstrates that the 
thermodynamic and the dynamical (in)stabilities of the system do not generically correlate. 
Since black holes with translationary invariant horizons in asymptotically anti-de-Sitter 
space-time are dual (albeit sometimes in a purely phenomenological way) to some strongly coupled 
field theory, one expects that it should be possible to construct 
a counter-example of generalized CSC as well.  In the next section we show that the 
generalized CSC\footnote{The violation of the 
canonical CSC in this system is shown in \cite{bpcsc}.} 
is violated for the exotic hairy black branes introduced in \cite{bp2}.

\section{Counter-example to generalized CSC in gravity}

Consider the following effective (3+1)-dimensional gravitational action \cite{bp2}:
\begin{equation}
\begin{split}
S_4=&\frac{1}{2\kappa^2}\int dx^4\sqrt{-\gamma}\left[R+6
-\frac 12 \left(\nabla\phi\right)^2+\phi^2-\frac 12 \left(\nabla\chi\right)^2-2\chi^2-g \phi^2 \chi^2
\right]\,,
\end{split}
\eqlabel{s4}
\end{equation}
where $g$ is a coupling constant\footnote{In numerical analysis we set $g=-100$.}.
This effective action admits asymptotically $AdS_4$ hairy black brane solutions with translationary invariant horizon. 
Specifically, the background geometry takes form 
\begin{equation}
ds_4^2=-c_1(r)^2\ dt^2+c_2(r)^2\ \left[dx_1^2+dx_2^2\right]+c_3^2\ dr^2\,,\qquad \phi=\phi(r)\,,\qquad 
\chi=\chi(r)\,.
\eqlabel{background}
\end{equation}
We find it convenient to introduce
a new radial coordinate $x$ as follows
\begin{equation}
1-x\equiv \frac{c_1(r)}{c_2(r)}\,,
\eqlabel{xdef}
\end{equation} 
so that $x\to 0$ corresponds to the boundary asymptotic, and $y\equiv 1-x\to 0$ 
corresponds to a regular Schwarzschild horizon asymptotic.
Further introducing 
\begin{equation}
c_2(x)=\frac{a(x)}{(2x-x^2)^{1/3}}\,,
\eqlabel{defa}
\end{equation}
the equations of motion obtained from \eqref{s4}, with the background 
ansatz \eqref{background}, imply 
\begin{equation}
\begin{split}
a=&\a\left(1-\frac{1}{40}\ p_1^2\ x^{2/3}-\frac{1}{18}\ p_1 p_2\ x+\calo(x^{4/3})\right)\,,\\
\phi=&p_1\ x^{1/3}+p_2\ x^{2/3}+\frac{3}{20} p_1^3 x+\calo(x^{4/3})\,,\\
\c=&\c_4\left(x^{4/3}+\left(\frac 17 g -\frac{3}{70}\right)p_1^2\ x^2+\calo(x^{7/3})\right)\,,
\end{split}
\eqlabel{boundary}
\end{equation}
near the boundary $x\to 0_+$, and 
\begin{equation}
\begin{split}
a=\a\left(a_0^{h}+a_1^h\ y^2+\calo(y^4)\right)\,,\qquad \phi=p_0^h+\calo(y^2)\,,\qquad  \c=c_0^h+\calo(y^2)\,,
\end{split}
\eqlabel{hor}
\end{equation}
near the horizon $y=1-x\to 0_+$. 
Apart from the overall scaling factor $\a$ (which is related to the temperature), the background is 
uniquely specified with 3 UV coefficients $\{p_1,p_2,\c_4\}$ and 4 IR coefficients 
$\{a_0^h,a_1^h,p_0^h,c_0^h\}$.  

It is straightforward to compute the temperature $T$ and the entropy density $s$ of the black brane solution 
\eqref{background}:
\begin{equation}
\left(\frac{8\pi T}{\a}\right)^2=\frac{6 (a^h_0)^3 (6-2 (\c^h_0)^2+(p^h_0)^2-g (p^h_0)^2 (c^h_0)^2)}{3 a^h_1+a^h_0}\,,
\eqlabel{tbh}
\end{equation}
\begin{equation}
\hat{s}\equiv \frac{384}{c}\ s= 4\pi\a^2\ (a_0^h)^2\,,
\eqlabel{sbh}
\end{equation}
where 
\begin{equation}
c=\frac{192}{\k^2}\,,
\eqlabel{c}
\end{equation} 
is the central charge of the UV fixed point. 
The free energy density $\calf$ and the energy density $\cale$  are given by 
\begin{equation}
\begin{split}
\hat{\calf}\equiv\frac{384}{c}\ \calf=&\a^3 \left(2-\frac 16  p_1 p_2-\frac {(a_0^h)^3}{2}
 \sqrt{\frac{6 a^h_0 (6-2 (c^h_0)^2+(p^h_0)^2-g (p^h_0)^2 (c^h_0)^2)}{3 a^h_1+a^h_0}}\right)\,,\\
\hat{\cale}\equiv\frac{384}{c}\ \cale=&\a^3 \left(2-\frac 16  p_1 p_2\right)\,.
\end{split}
\eqlabel{p1fixed}
\end{equation} 
Lastly, we identify $\Lambda$,
\begin{equation}
\Lambda\equiv p_1\ \a\,,
\eqlabel{defl} 
\end{equation}
with the mass scale of the dual (boundary) field theory. Notice that the scalar field $\c$ can not have a non-zero 
non-normalizable coefficient as the latter would destroy the asymptotic $AdS_4$ geometry --- near the boundary, the non-normalizable mode of $\c$ behaves as\footnote{Further details of the hairy 
black brane solutions can be found in \cite{bp2}.} $\c\sim x^{-1/3}$.   

For a given set of $\{\a,p_1\}$ there is a discrete set of the remaining parameters 
\[
\{p_2,\ \c_4,\ a_0^h,\ a_1^h,\ p_0^h,\ c_0^h\}
\]
characterizing black brane solutions. One of these solutions has $\{\c_4,\ c_0^h\}=\{0,0\}$
and describes the black brane without the condensate of the $\c$ field. All the other solutions 
have  $\{\c_4,\ c_0^h\}\ne \{0,0\}$ and describe the "exotic black branes'' \cite{bp2}.
In was shown in \cite{bpcsc} that all the exotic black branes contain a tachyonic quasinormal mode, 
and thus are dynamically unstable. In the remainder of this section we show that 
exotic black  branes are not only thermodynamically stable in a canonical way \cite{bp2}, they are thermodynamically 
stable in a generalized way as well,  with $\Lambda$ being treated as a generalized charge.

\begin{figure}[t]
\begin{center}
\psfrag{er}{{$\frac{\hat{\cale}}{\hat{s}^{3/2}}$}}
\psfrag{mu}{{$\frac{\Lambda}{\hat{s}^{1/2}}$}}
\includegraphics[width=5in]{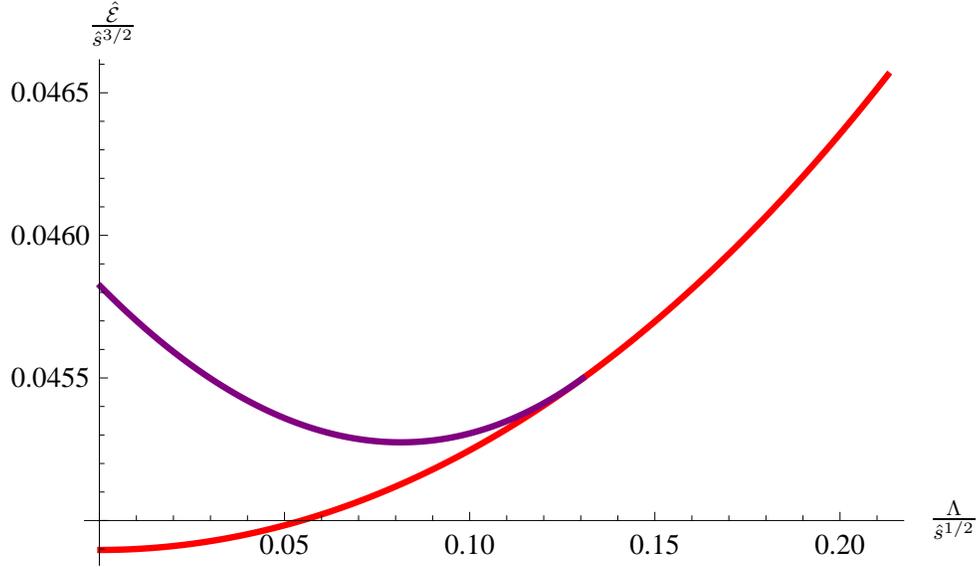}
\end{center}
  \caption{(Colour online)
The energy density of the black branes with the scalar condensate (purple points) and without 
the scalar condensate (red points).
 } \label{figure1}
\end{figure}

Given a dataset $\{p_1,\ p_2,\ \c_4,\ a_0^h,\ a_1^h,\ p_0^h,\ c_0^h\}$ for each of the discrete branches of the 
black brane solutions we can construct parametric dependence of $\frac{\hat{\cale}}{\hat{s}^{3/2}}$ versus 
$\frac{\Lambda}{\hat{s}^{1/2}}$, \ie the function $(x,\calg(x))$ such that 
\begin{equation}
\hat{\cale}={\hat{s}^{3/2}}\ \calg\left(\frac{\Lambda}{\hat{s}^{1/2}}\right)\,.
\eqlabel{enl}
\end{equation}
Given \eqref{tbh}-\eqref{defl} we have 
\begin{equation}
\frac{\Lambda}{\hat{s}^{1/2}}=\frac{p_1}{2\pi^{1/2}a_0^h}\,,
\eqlabel{xg}
\end{equation}
\begin{equation}
\frac{\hat{\cale}}{\hat{s}^{3/2}}=\frac{12-p_1 p_2}{48\pi^{3/2}(a_0^h)^3}\,.
\eqlabel{yg}
\end{equation}
Figure~\ref{figure1} presents the function $(x,\calg(x))$ for the black branes without 
the condensate of the $\c$ scalar (the red points), and with the condensate of the $\c$ 
scalar (purple points).  

\begin{figure}[t]
\begin{center}
\psfrag{detH}{{$\det\left(\hhh^{\hat{\cale}}_{\hat{s},\Lambda}\right)$}}
\psfrag{del2}{{$\hat{s}^{1/2}\ \frac{\del^2 \hat{\cale}}{\del\hat{s}^2}$}}
\psfrag{mu}{{$\frac{\Lambda}{\hat{s}^{1/2}}$}}
\includegraphics[width=3in]{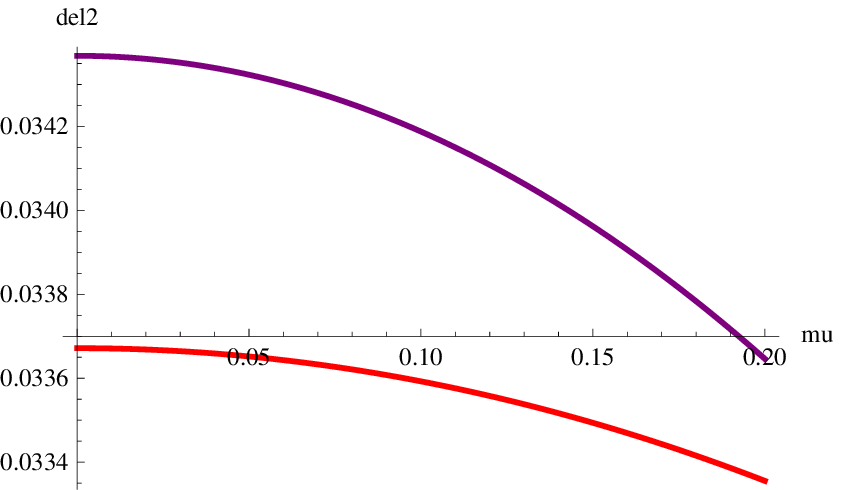}
\includegraphics[width=3in]{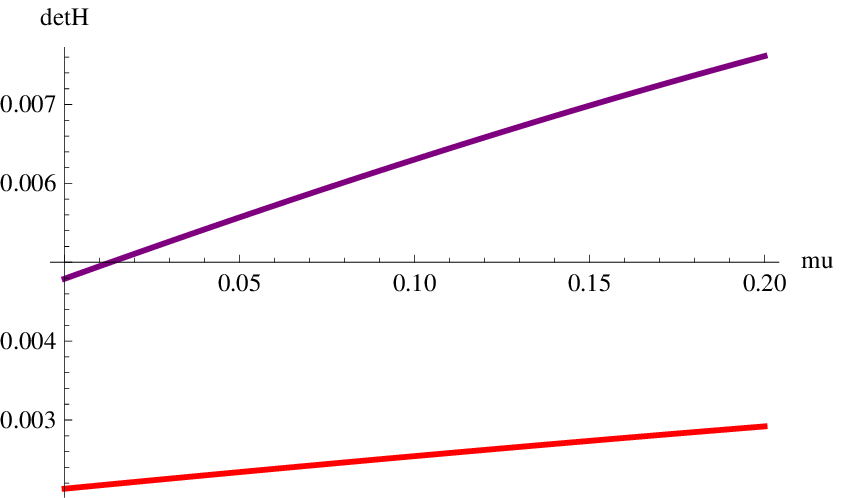}
\end{center}
  \caption{(Colour online)
Canonical $\hat{s}^{1/2}\ \frac{\del^2 \hat{\cale}}{\del\hat{s}^2}>0$
(left plot) and the generalized $\det\left(\hhh^{\hat{\cale}}_{\hat{s},\Lambda}\right)>0$
(right plot) thermodynamic stability criteria 
for the dynamically stable (red curves) and the dynamically unstable (purple curves)
hairy black branes.  
 } \label{figure2}
\end{figure}

The following fits to $\calg(x)^{red}$ and $\calg(x)^{purple}$ are indistinguishable with a naked eye from the 
data points in Figure~\ref{figure1}:
\begin{equation}
\begin{split}
\calg(x)^{red}=&0.0448955 + 0.000128216\ x + 0.0316168\ x^2 + 0.0212735\ x^3\,,\\
\calg(x)^{purple}=&0.0458244 - 0.0130892\ x + 0.0721953\ x^2 + 0.06829850585\ x^3\,.
\end{split}
\eqlabel{functions}
\end{equation}

We are now ready to analyze the canonical and the generalized thermodynamic stability criterion for the 
hairy black branes.
\nxt In the canonical case we require that the Hessian
\begin{equation}
\hhh^{\hat{\cale}}_{\hat{s}}
\eqlabel{can1}
\end{equation}
be positive, 
which translates into 
\begin{equation}
0\ <\ \hat{s}^{1/2}\ \frac{\del^2 \hat{\cale}}{\del\hat{s}^2}=\biggl\{\frac 34\ \calg(x)-\frac 34\ x\ \calg'(x)
+\frac 14 x^2\ \calg''(x)\biggr\}\bigg|_{x=\frac{\Lambda}{\hat{s}^{1/2}}}\,.
\eqlabel{first}
\end{equation}
\nxt In the generalized case, the scale $\Lambda$ is treated as one of the  charges $Q_A$; thus, the 
thermodynamic stability criterion becomes
the positivity of the Hessian
\begin{equation}
\hhh^{\hat{\cale}}_{\hat{s},\Lambda}\,,
\eqlabel{can2}
\end{equation}
which in addition to \eqref{first} requires that  
\begin{equation}
0\ <\ \det\left(\hhh^{\hat{\cale}}_{\hat{s},\Lambda}\right)=
\biggl\{
\frac 34\ \calg''(x)\ \calg(x)+
\frac 14 x\ \calg''(x)\ \calg'(x)-\left(\calg'(x)
\right)^2
\biggr\}\bigg|_{x=\frac{\Lambda}{\hat{s}^{1/2}}}\,.
\eqlabel{first22}
\end{equation}

The results of the stability analysis \eqref{first} and \eqref{first22} are presented in Figure~\ref{figure2}.
Much like in the simple statistical model of section 2 both the canonical and the generalized thermodynamic 
stability criteria characterize the hairy black branes (with or without the scalar condensate $\c$) 
as being stable. As established in \cite{bpcsc}, the hairy black branes with the non-zero condensate of 
$\c$ are dynamically unstable. Thus, we conclude that generalizing the thermodynamic stability criterion to include 
(in an appropriate manner) the asymptotic coefficients of scalar fields sourcing the 
black branes in an asymptotically anti-de-Sitter space-time can not validate the ``Correlated Stability Conjecture''.
 
Much like in the statistical model in section 2, it is clear that, once one is sufficiently 
close to the transition (so that the tachyon condensate contribution to the thermodynamics is negligible), 
any redefinition of the generalized charge $Q_A= f(\Lambda)$ would change the 
thermodynamic stability of both classically stable and unstable phases in identical manner.  
Thus, insisting that the classically stable phase is thermodynamically stable (in a generalized way) as well
would imply that the classically unstable phase is also thermodynamically stable.
From this perspective our counter-example of the CSC conjecture is robust with respect to a definition of the 
generalized charge\footnote{As emphasized in section 2 we assume that $Q_A$ depends only on the microscopic 
scale(s) of the theory. } $Q_A$.

\section*{Acknowledgments}
We would like to thank Barak Kol for commenting on \cite{bpcsc}
that prompted us to perform analysis reported here.
Research at Perimeter Institute is
supported by the Government of Canada through Industry Canada and by
the Province of Ontario through the Ministry of Research \&
Innovation. AB gratefully acknowledges further support by an NSERC
Discovery grant and support through the Early Researcher Award
program by the Province of Ontario.


\begin{thebibliography}{99}

\bibitem{gm1}
  S.~S.~Gubser and I.~Mitra,
  ``Instability of charged black holes in anti-de Sitter space,''
  arXiv:hep-th/0009126.


\bibitem{gm2}
  S.~S.~Gubser and I.~Mitra,
  JHEP {\bf 0108}, 018 (2001)
  [arXiv:hep-th/0011127].


\bibitem{b}
  A.~Buchel,
  Nucl.\ Phys.\  B {\bf 731}, 109 (2005)
  [arXiv:hep-th/0507275].


\bibitem{ccsc}
  J.~J.~Friess, S.~S.~Gubser and I.~Mitra,
  Phys.\ Rev.\  D {\bf 72}, 104019 (2005)
  [arXiv:hep-th/0508220].

\bibitem{bpcsc}
  A.~Buchel and C.~Pagnutti,
  Phys.\ Lett.\  B {\bf 697}, 168 (2011)
  [arXiv:1010.5748 [hep-th]].


\bibitem{m9711}
J.~M.~Maldacena,
Adv.\ Theor.\ Math.\ Phys.\  {\bf 2}, 231 (1998)
[Int.\ J.\ Theor.\ Phys.\  {\bf 38}, 1113 (1999)]
[arXiv:hep-th/9711200].



\bibitem{bp2}
  A.~Buchel and C.~Pagnutti,
  Nucl.\ Phys.\  B {\bf 824}, 85 (2010)
  [arXiv:0904.1716 [hep-th]].


\bibitem{bp3}
  A.~Buchel and C.~Pagnutti,
  Nucl.\ Phys.\  B {\bf 834}, 222 (2010)
  [arXiv:0912.3212 [hep-th]].



\end{thebibliography}
\end{document}